\documentclass[prd,preprintnumbers,floatfix,
nofootinbib,superscriptaddress]{revtex4}

\usepackage{float}
\usepackage{nicefrac}
\usepackage{mathtools}
\usepackage{amsfonts} 
\usepackage{amssymb} 
\usepackage{amsmath} 
\usepackage{bbold}
\usepackage{bm} 
\usepackage{graphicx} 
\usepackage{subfigure} 
\usepackage{array} 
\usepackage{dcolumn} 
\usepackage{bm} 
\usepackage{latexsym} 
\usepackage{longtable} 
\usepackage{hyperref} 
\usepackage{verbatim}
\usepackage{epsfig}
\usepackage{slashed}
\usepackage{color}

\usepackage{glossaries-extra}
\setabbreviationstyle[acronym]{long-short}
\glssetcategoryattribute{acronym}{nohyperfirst}{true}

\newcommand{\cA}[0]{\mathcal A}

\newcommand{\cD}[0]{\mathcal D}

\newcommand{\cL}[0]{\mathcal L}

\newcommand{\cO}[0]{\mathcal O}

\newcommand{\cR}[0]{\mathcal R}
\newcommand{\cS}[0]{\mathcal S}
\newcommand{\cT}[0]{\mathcal T}

\newcommand{\wt}[0]{\widetilde}

\newcommand{\thr}[0]{{\rm th}}

\newcommand{\chpt}{$\chi$PT}
\newcommand{\SU}{{\rm SU}}

\newcommand{\Luscher}[0]{Luscher:1986n2,Luscher:1991n1}

\newcommand{\KSS}[0]{Kim:2005gf}
\newacronym{CMF}{CMF}{center-of-momentum frame}

\begin{document}


\title{Applicability of the two-particle quantization condition to partially-quenched
theories}

\author{Zachary T. Draper}
\email[e-mail: ]{ztd@uw.edu}
\affiliation{Physics Department, University of Washington, Seattle, WA 98195-1560, USA}

\author{Stephen R. Sharpe}
\email[e-mail: ]{srsharpe@uw.edu}
\affiliation{Physics Department, University of Washington, Seattle, WA 98195-1560, USA}


\date{\today}

\begin{abstract}
Partial quenching allows one to consider correlation functions and amplitudes
that do not arise in the corresponding unquenched theory.
For example, physical $s$-wave pion scattering can be decomposed into $I=0$ and $2$ amplitudes, 
while, in a partially-quenched extension, the larger symmetry group
implies that there are more than two independent scattering amplitudes.
It has been proposed that the finite-volume quantization condition of L\"uscher 
holds for the correlation functions associated with each of the two-particle amplitudes
that arise in partially-quenched theories.
Using partially-quenched chiral perturbation theory,
we show that this proposal fails for those correlation functions
for which the corresponding one-loop amplitudes do not satisfy $s$-wave unitarity.
For partially-quenched amplitudes that, while being unphysical, do satisfy one-loop $s$-wave unitarity,
we argue that the proposal is plausible. Implications for previous work are discussed.
 \end{abstract}


\nopagebreak

\maketitle


\section{Introduction}
\label{sec:intro}


%

Simulations of lattice QCD (LQCD) and related theories
 naturally separate into the generation of gauge fields
and the calculation of quark propagators on these gauge fields.
This separation allows one the freedom to ``partially quench'', which in broadest terms means
treating valence and sea quarks differently. 
This can be done by using different valence- and sea-quark masses, or, as we consider here,
by keeping valence- and sea-quark masses the same but 
considering correlation functions---combinations of quark propagators---that cannot arise in QCD.
In either case, one obtains extra information about the underlying theory.
The important questions are then whether that information is useful, e.g. for extracting physical
quantities, and, if so, how the desired output is extracted.

There are many proposals for the extraction of additional information using
partially-quenched (PQ) theories, several of which have been applied in practice.
 Most involve the use of partially-quenched chiral perturbation
theory (PQ\chpt), which was developed in Ref.~\cite{Bernard:1993sv}.
For reviews see Refs.~\cite{Sharpe:2006pu,Golterman:2009kw}.
One usage is to aid the chiral extrapolation/interpolation to the physical point using a range
of valence quark masses~\cite{Sharpe:2000bc}, and this has been widely applied in practice.
Another example, closely related to the discussion below,
is the proposal of Ref.~\cite{Hansen:2011mc} to separately calculate contributions
to $\pi^+\pi^+$ correlators with and without quark exchange, so as to
determine the lattice-spacing-dependent low-energy coefficients (LECs) of ``Wilson\chpt'' 
(i.e. \chpt\ applied to Wilson fermions).
One can also use PQ\chpt\ to estimate quark-disconnected diagrams from more easily
calculated correlators, e.g. in the hadronic vacuum polarization contribution to muonic $g-2$~\cite{DellaMorte:2010aq}, 
or in pion scattering~\cite{Acharya:2017zje}.
We also note the use of partially quenched theories to study the spectral density
of the Dirac operator~\cite{Osborn:1998qb,DelDebbio:2005qa,Sharpe:2006ia,Damgaard:1998xy}
and its relation to random matrix theories~\cite{Damgaard:2010cz}.

This paper considers the proposals of Refs.~\cite{Acharya:2019meo,KalmahalliGuruswamy:2020uxi},
which concern, respectively, $\pi\pi$ and $K \pi$ scattering in QCD.
These works consider PQ extensions of QCD containing additional valence quarks (and
corresponding ghost quarks) such that the flavor symmetry group extends from 
$\SU(2)$ to $\SU(4|2)$ for $\pi\pi$ scattering,
and from $\SU(3)$ to $\SU(4|1)$ for $K\pi$ scattering.\footnote{%
The notation here is that the $\SU(n|m)$ symmetry corresponds to $n-m$ sea quarks,
$m$ valence quarks, and $m$ (commuting) ghost quarks. 
The flavor symmetry of the unquenched subsector is $\SU(n-m)$.}
Within these extensions the scattering amplitudes of pions and kaons (composed of
either sea or valence quarks) can be decomposed into irreducible representations (irreps)
of the quark flavor symmetry group, which is $\SU(4)$ in both cases.
This group is larger than the symmetry group of the unquenched sea quarks,
which is $\SU(2)$ and $\SU(3)$, respectively, for the two cases.
This implies that the number of irreps, and associated independent amplitudes,
is larger in the PQ theory than in the unquenched theory.
For example, for $s$-wave $\pi\pi$ scattering, there are four independent amplitudes
in the quark sector of the PQ theory, 
corresponding to the symmetric product of two adjoints in $\SU(4)$~\cite{Sharpe:1992pp},
\begin{equation}
(\bm {15} \times \bm {15})_{\rm sym} = \bm 1 + \bm{ 15} + \bm {20} + \bm {84}\,,
\label{eq:SU4decomp}
\end{equation}
whereas for unquenched pions in QCD the decomposition contains only
two amplitudes, those with isospin 0 and 2.
The physical $I=2$ amplitude is equivalent to the $\bm{84}$, while the $I=0$ amplitude is
equivalent to that obtained from an operator that is a linear combination of the $\bm{1}$ 
with elements of the $\bm{15}$, $\bm{20}$ and $\bm{84}$.\footnote{%
The linear combination is $\tfrac15 \bm{1}+\tfrac12 \bm{15} + \tfrac14 \bm{20} + \tfrac1{20} \bm{84}$,
as can be seen by considering the quark contractions.
That this leads to a physical amplitude is not immediately clear from the group theory,
but follows from the fact that operators living entirely in the unquenched sector are free
from PQ artifacts.}
The remaining independent amplitudes, 
which we take to be those in the $\bm{15}$ and $\bm{20}$ irreps, 
are artifacts of the PQ theory,
and correspond to linear combinations of Wick contractions
that are not present in the unquenched theory.

We focus in this paper on nonsinglet irreps of the quark flavor group,
as this simplifies the calculations and yet allows us to make our main points.
We make some brief comments on the singlet irrep in the conclusions.

A central claim of Refs.~\cite{Acharya:2019meo,KalmahalliGuruswamy:2020uxi} is the following:
for each of the PQ amplitudes, 
the corresponding PQ two-particle correlation function
determines a spectrum of energies that can be related to the scattering amplitude using
the two-particle quantization condition of L\"uscher~\cite{\Luscher}.
We stress that this claim consists of two parts, the first that the correlation function
can be used to extract a spectrum of states in the usual way, and the second that
the resulting spectrum can be analyzed using the two-particle quantization condition.
To make these statements concrete, consider the $r=\bm{15}$ irrep in PQ pion scattering,
and let $\cO_r$ be an operator with these quantum numbers.
The first part of the claim is that
\begin{equation}
\langle \cO_r(\tau) \cO_r^\dagger(0) \rangle = \sum_{n=0}^{\infty} c_n e^{-E_n \tau}
\label{eq:2pt}
\end{equation}
where the expectation value is in the underlying PQ theory,
$\tau$ is  Euclidean time, the coefficients $c_n$ are real and positive, 
and the energies $E_n$ are real and bounded below.
For convenience we order them $E_0 \le E_1 \le E_2$, etc.
The second part of the claim is that we can insert the resulting
energies $E_n$ into L\"uscher's quantization condition and correctly obtain the
scattering amplitude in the corresponding irrep.
We know that this procedure is valid for the $\bm{84}$ irrep, as it
corresponds to a physical correlation function in QCD.
The issue is whether it remains valid for the irreps available only in the PQ theory.

If correct, this proposal is quite powerful, as, combined with PQ\chpt, 
it allows the determination of results from types of contractions
that are difficult to calculate numerically from those that are simpler to 
evaluate~\cite{Acharya:2017zje,Acharya:2019meo,KalmahalliGuruswamy:2020uxi,
Guruswamy:2021nph}.
It is, however, {\em prima facie} surprising, since both the existence of a normal spectral
decomposition and the derivation of the two-particle quantization condition rely on unitarity,
which the PQ theory violates~\cite{Bernard:1993sv,Bernard:2013kwa}.

In this work we introduce a necessary (but not sufficient) criterion for whether this proposal
holds in a given PQ irrep. We apply it in detail to $\pi\pi$ scattering, which allows us
to develop alternative criteria for when the proposal will fail.
We then apply these criteria to other systems. 

The remainder of this paper is organized as follows.
Section~\ref{sec:setup} describes the setup within PQ\chpt\ that we use to
formulate our initial criterion.
Section~\ref{sec:calc} presents the results of the application of
this criterion to $\pi\pi$ scattering.
Section~\ref{sec:gen} interprets these results using quark-line diagrams,
presents two alternative versions of the criterion, and discusses
the extent to which our necessary criterion is actually sufficient.
Section~\ref{sec:implications} discusses a few applications of our criteria,
and describes their implications for previous work.
Finally, 
Sec.~\ref{sec:conc} presents some closing comments.

\section{Theoretical setup}
\label{sec:setup}

We follow closely the methodology of Ref.~\cite{Hansen:2011mc}, although here we
work in the continuum.
We introduce two valence quarks in addition to the physical up and down quark,
as well as two ghost quarks,
with all quarks and ghost quarks having a common mass $m$.
Thus the graded flavor symmetry group is $\SU(4|2)$.\footnote{%
Superficially, this differs from the setup in Ref.~\cite{Hansen:2011mc}, where 
four valence quarks and corresponding ghost quarks were introduced, 
leading to an $\SU(6|4)$ flavor symmetry.
For the purposes of our calculations, however, the difference is trivial: both calculations build
the required operators out of four degenerate quark fields, 
leading to identical quark-level contractions and \chpt\ results.
The additional valence quarks were introduced in Ref.~\cite{Hansen:2011mc} 
in order to calculate a different quantity, as described in Appendix B of that work.}
This is the same theoretical setup as used in Refs.~\cite{Acharya:2017zje,Acharya:2019meo}.
We label the quark fields $q_i$, $i=1-4$, with $q_1=u$ and $q_2=d$ being 
thought of as the sea quarks, although this choice is purely conventional given the
exact $\SU(4)$ quark flavor symmetry.
The two ghost quarks are labeled $\wt q_5$ and $\wt q_6$.
Using this collection of fields we can write down the PQQCD functional integral
in Euclidean space as described in Ref.~\cite{Bernard:1993sv}.

The long-distance properties of this theory are described by PQ\chpt.
The degrees of freedom of this effective theory are the
pseudo-Goldstone (PG) boson and fermion fields, 
which are collected into the following straceless\footnote{%
We use ``straceless'' as a shorthand for having a vanishing supertrace,
and refer to the latter as ``strace'' or ``str".}
$6\times6$ matrix~\cite{Bernard:1993sv,Sharpe:2001fh},
\begin{equation}
\Pi = \frac1{\sqrt2} \begin{pmatrix}
\tfrac1{\sqrt2}\pi_0\!+\!\tfrac12\eta_4 \!+\!\tfrac12\phi_1
 & \pi_{12} & \pi_{13} & \pi_{14}  & \omega_{15} & \omega_{16} \\
\pi_{21} & -\tfrac1{\sqrt2}\pi_0\!+\!\tfrac12\eta_4 \!+\!\tfrac12\phi_1
 & \pi_{23} & \pi_{24} & \omega_{25} & \omega_{26} \\
\pi_{31} & \pi_{32} &\tfrac1{\sqrt2} \pi'_0\!-\!\tfrac12\eta_4 \!+\!\tfrac12\phi_1 
 & \pi_{34} & \omega_{35} & \omega_{36} \\
\pi_{41} & \pi_{42} & \pi_{43} & -\tfrac1{\sqrt2}\pi'_0\!-\!\tfrac12\eta_4 \!+\!\tfrac12 \phi_1
 & \omega_{45} & \omega_{46} \\
\omega_{51} & \omega_{52} &\omega_{53} &\omega_{54} & \tfrac1{\sqrt2}\phi_0 \!+\! \phi_1 
 &\phi_{56} \\
\omega_{61} & \omega_{62} &\omega_{63} &\omega_{64} & \phi_{65} 
 & -\tfrac1{\sqrt2} \phi_0 \!+\! \phi_1
\end{pmatrix}\,.
\label{eq:Pi}
\end{equation}
Here $\pi_0$ is the usual neutral pion field (in isosymmetric QCD), with $\pi'_0$ being the
corresponding valence field, while $\eta_4$ is the four flavor generalization of
the $\eta$ meson in QCD. 
The 15 fields in the quark sector of  $\Pi$ fill out the $\SU(4)$ adjoint irrep of PG bosons (PGBs).
The $\omega_{ij}$ are PG fermion (PGF) fields corresponding to quark-ghost combinations,
while the $\phi_0$, $\phi_{ij}$, and $\phi_1$ are PG ghost bosons (with propagators having
an unphysical overall sign).

We will only need the leading order (LO) chiral Lagrangian for this theory, which is given
in Euclidean space by
\begin{equation}
\cL_{\rm LO} = \frac{f^2}4 {\rm str}\!\left(\partial_\mu \Sigma^\dagger \partial_\mu \Sigma \right)
-\frac{m B_0 f^2}{2} {\rm str}\!\left(\Sigma^\dagger+\Sigma \right)
\,,
\label{eq:LchptLO}
\end{equation}
where $f$ and $B_0$ are the usual LO low-energy coefficients (LECs),
with the former defined in the small $f$ convention,
and
\begin{equation}
\Sigma = \exp( 2 i\Pi/f)\,.
\end{equation}
The masses of all the PGBs and PGFs are given, at LO, by
\begin{equation}
M_\pi^2 = 2 B_0 m\,.
\end{equation}
We note also that, because all the quarks and ghost quarks are degenerate,
the PG propagators all contain only single poles, with no double-pole 
contributions~\cite{Sharpe:2000bc,Sharpe:2001fh}.

The relation of the pion fields in \chpt\ to operators in the underlying theory can be
worked out using the spurion method. The resulting correspondence
is given for $j,k=1\!-\!4$ by (see, e.g. Ref.~\cite{Hansen:2011mc})
$\bar q_j \gamma_5 q_k(x) = c (\Sigma-\Sigma^\dagger)_{kj}$, 
with $c$ a known constant that will cancel in the ratios considered below.
Choosing $j\ne k$ for simplicity, the right-hand side of this relation is proportional to $\pi_{kj}$,
up to chiral corrections proportional to $\pi^2/f^2$.
As discussed in Sec.~C.1 of Ref.~\cite{Hansen:2011mc}, these correction terms lead
to subleading contributions to the correlation functions that lie beyond the accuracy that we
consider. Thus we effectively have the correspondence $\pi_{kj} \sim \bar q_j\gamma_5 q_k$,
and, similarly, $\pi^0 \sim \tfrac1{\sqrt2}(\bar q_1 \gamma_5 q_1 -\bar q_2 \gamma_5 q_2)$
and $\eta_4 \sim\tfrac12 (\bar q_1 \gamma_5 q_1 + \bar q_2 \gamma_5 q_2 - \bar q_3\gamma_5 q_3
-\bar q_4 \gamma_5 q_4)$, with a common constant of proportionality.
Using this correspondence, the definitions given below apply both in the
underlying theory, PQQCD, and in the effective theory, PQ\chpt, and we will move 
between these two representations as needed.

The quantities that we calculate are finite-volume correlation functions involving operators coupling to pairs of PGB fields.
These are built from fields having zero spatial momentum, e.g.
\begin{equation}
\wt\pi_{12}(\tau) = \int_L d^3x\, \pi_{12}(x,\tau)\,,
\end{equation}
where the subscript $L$ indicates that the integral is over a cubic box of side $L$.
We assume periodic boundary conditions (PBC) in spatial directions on the PGBs and PGFs, 
which follow if the quark fields satisfy PBC, as it standard in LQCD simulations.
We take the Euclidean time extent to be infinite.
We then construct two-PGB operators by forming linear combinations of the building blocks
\begin{equation}
\wt \pi_{ij}(\tau) \wt \pi_{k\ell} (\tau)\,,
\end{equation}
chosen to have definite flavor quantum numbers. 
Explicit examples of such operators are given in the following section.

In a physical, unitary theory, we can describe the properties of these operators as follows.
In the absence of interactions, such operators would simply couple to two pions at rest.
In the presence of interactions, however, they couple to all states 
with vanishing total three-momentum, $\bm P=0$, that have the same flavor as the operator.
In particular, these operators
couple to the lightest two-PGB state of the chosen flavor.
We refer to this as the threshold state, with energy $E_0$.
Of particular interest is the energy shift $\delta E_0= E_0 - 2 M_\pi$, 
for this is given by the threshold expansion of the 
L\"uscher quantization condition~\cite{Luscher:1986n2}
\begin{equation}
\delta E_0 = -\frac{\cA_\thr}{4 M_\pi^2 L^3} \left[1 + 
c_1 \frac{\cA_\thr}{16\pi M_\pi L} + \cO(L^{-2}) \right]
\,.
\label{eq:LuscherTH}
\end{equation}
Here $\cA_\thr$ is the scattering amplitude at threshold
in the flavor channel under consideration,
which is related to the scattering length $a_0$ by\footnote{%
Here we are using the normalization appropriate for distinguishable particles.}
\begin{equation}
\cA_{\rm th}= 16 \pi M_\pi a_0\,,
\label{eq:a0toAth}
\end{equation}
while $c_1\approx -2.84$ is a known geometric constant.
We stress that both $\cA_\thr$ and $a_0$ are infinite-volume quantities.
Higher-order terms in $1/L$ in Eq.~(\ref{eq:LuscherTH}) are known,
but will not be needed here.
There are also (in general unknown) corrections to Eq.~(\ref{eq:LuscherTH})
that are exponentially suppressed in $M_\pi L$.
In our theoretical calculation these can always be made arbitrarily small, compared to the
power-law dependence that we control, by taking $L$ large enough, 
and henceforth we ignore them.

To obtain $\delta E_0$ from finite-volume correlators, it is convenient to use ratios of
the two-PGB correlator to the square of single-particle correlators. Let $\cO_r(\tau)$ denote
(as in the Introduction) a two-PGB operator transforming in the flavor irrep $r$.
Then we consider 
\begin{equation}
R_r(\tau) =\frac{
\langle \cO_r(\tau) \cO_r^\dagger(0) \rangle}
{\langle \wt\pi_{12}(\tau) \wt \pi_{21}(0) \rangle^2}\,.
\label{eq:Rr}
\end{equation}
where the choice of flavor of the single-PGB correlator in the denominator is arbitrary, 
because all such correlators fall asymptotically as $\exp(-M_\pi |\tau|)$,
due to the $\SU(4)$ flavor symmetry.
For a physical theory, the ratio behaves at large $|\tau|$ as
\begin{equation}
R_r(\tau) = Z_r e^{-\delta E_{0,r} |\tau|} + \textrm{excited-state contributions}\,,
\label{eq:Rrasymp}
\end{equation}
where $\delta E_{0,r}$ is the threshold energy shift in the given flavor irrep, 
and $Z_r$ is a positive real constant.
In an analytic calculation, the contribution from excited states can be separated
by hand, since they have a distinctive exponential falloff,
with $\delta E_i \propto 1/L^2$ asymptotically.
The first part of our criterion is then to check that what remains after excited-state contributions
have been dropped is an exponential. More precisely, we test that the ratio contains
the first three terms in the expansion,
\begin{equation}
R_r(\tau) = Z_r \left[1 - |\tau| \delta E_{0,r} + \frac{\tau^2}2 (\delta E_{0,r})^2\right] + \dots\,, 
\label{eq:test1}
\end{equation}
where the ellipsis represents higher-order terms in $\tau$ as well as the excited-state contributions.

\begin{figure}[tb]
\begin{center}
\vspace{-12pt}
\includegraphics[width=\textwidth]{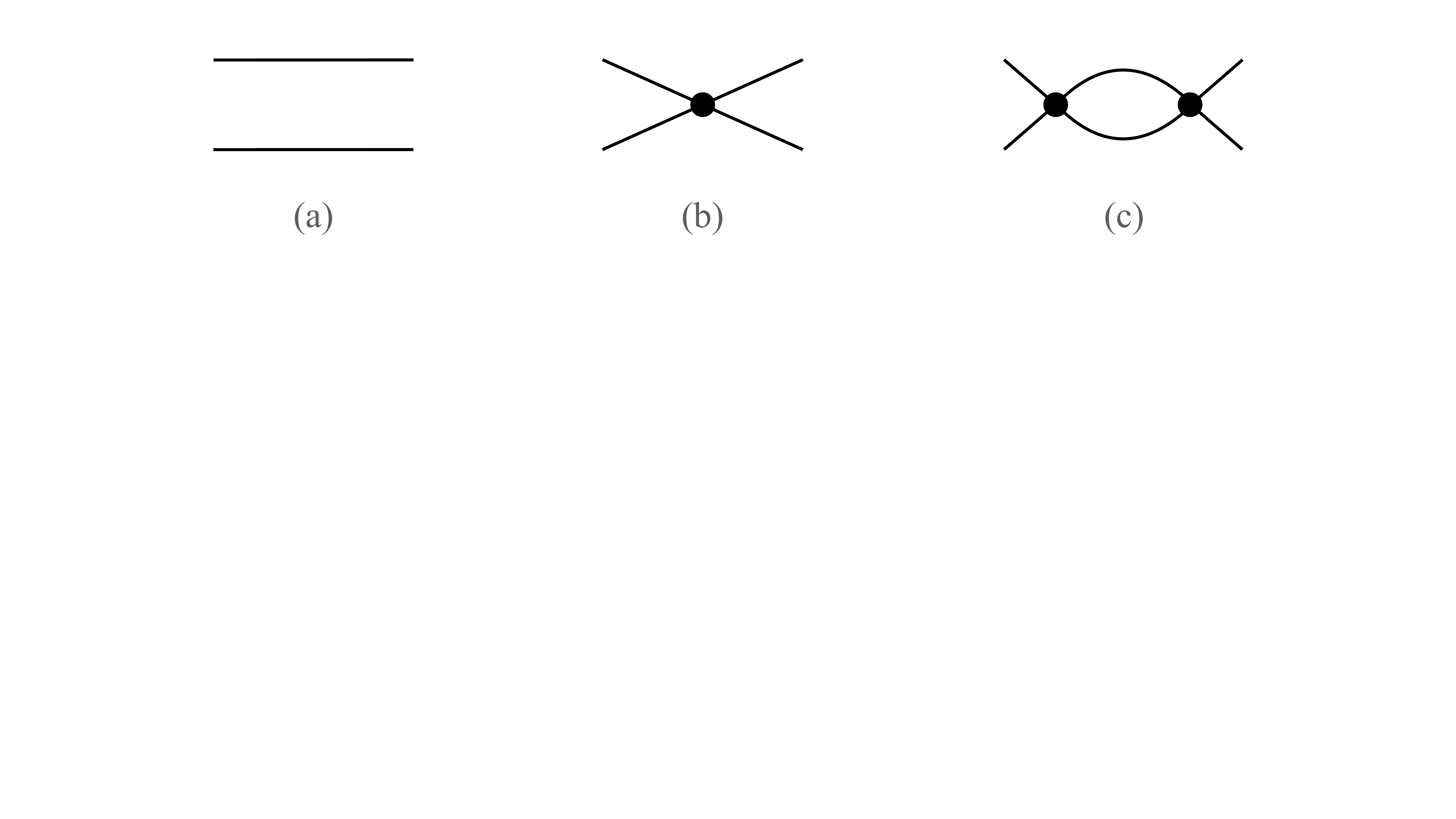}
\vspace{-2.9truein}
\caption{PQ\chpt\ diagrams contributing to $R_r(\tau)$, Eq.~(\ref{eq:Rr}).
Time runs horizontally. Solid lines are propagators of PGBs and PGFs, and
filled circles represent vertices obtained from the expansion of $\cL_{\rm LO}$,
Eq.~\ref{eq:LchptLO}. These are the diagrams needed to implement the
test discussed in the text.
\label{fig:chpt}
}
\end{center}
\end{figure}

The second part of our criterion tests whether the threshold expansion,
Eq.~(\ref{eq:LuscherTH}), holds to first nontrivial order in $1/L$.
It turns out that one can test this, as well as Eq.~(\ref{eq:test1}), by carrying out
a next-to-leading-order (NLO) calculation of the correlator ratio in \chpt.
To see how this works, we need several results that follow from
the similar calculation carried out in Ref.~\cite{Hansen:2011mc}.
The relevant diagrams are those shown in Fig.~\ref{fig:chpt}.
Diagram (a) gives the leading, $\tau$-independent term in the exponential in Eq.~(\ref{eq:test1}),
and one can choose the normalization of $\cO_r$ such that $Z_r=1$ at this stage.
Diagram (b) leads to a term linear in $|\tau|$, in which
$\delta E_{0,r}= - \cA_{\thr,r}^{\rm LO}/(4 M_\pi^2 L^3)$, which is the form expected at this order
from combining Eqs.~(\ref{eq:LuscherTH}) and (\ref{eq:test1}).
This result will hold for all irreps $r$, since the
diagrams contributing to the finite-volume correlator and the scattering amplitude are
essentially the same.
Diagram (b) also leads to corrections to $Z_r$ that are proportional to $1/(f^2 M_\pi L^3)$.

The key diagram for our test is that of Fig.~\ref{fig:chpt}(c).
In particular, for a physical theory, it leads both to the $\tau^2$ term in Eq.~(\ref{eq:test1})
with the LO form for the energy shift, $\delta E_{0,r}= - \cA_{\thr,r}^{\rm LO}/(4 M_\pi^2 L^3)$,
{\em and} to the first $1/L^4$ correction linear in $|\tau|$,
i.e.
\begin{equation}
\delta E_{0,r} \supset - c_1 \frac{(\cA_{\thr,r}^{\rm LO})^2}{64\pi M_\pi^3 L^4}\,.
\label{eq:c1term}
\end{equation}
It is these contributions that will turn out to have the wrong form for some irreps in the
PQ theory.

Diagram (c) also contributes to the leading $1/L^3$ part of the coefficient of $|\tau|$.
Together with diagrams containing NLO vertices, as well as self-energy diagrams, this
leads to the expected form of the linear term at this order~\cite{Hansen:2011mc}:
\begin{equation}
R_r(\tau) \supset |\tau| \frac{\cA_{\thr,r}^{\rm NLO}}{4 M_\pi^2 L^3}\,,
\end{equation}
where $\cA_{\thr,r}^{\rm NLO}$ is the complete threshold amplitude through NLO.
As at LO, this result holds for all irreps $r$, because the diagrams that contribute to
the $|\tau|/L^3$ term are the same as those leading to infinite-volume scattering.
The only difference is that the former involve momentum sums, while the latter
contain integrals, but the difference can be shown to be subleading,
behaving as $|\tau|/L^4$, and leads to the $c_1$ term in
Eq.~(\ref{eq:c1term})~\cite{Hansen:2011mc}.
As stressed in Ref.~\cite{Hansen:2011mc}, the result that the coefficient
of $|\tau|/L^3$ gives the scattering amplitude is a finite volume version of
the Lehman-Symanzik-Zimmermann reduction theorem.

Pulling all this together, we now state our criterion for whether a particular PQ channel can
be treated as if it were physical.
\begin{quote}
C1: {\em 
In order for a channel to
be described by the two-particle quantization condition,
one must find the following result\footnote{%
Here we are assuming the above-described normalization choice for $\cO_r$.}
if one calculates the diagrams of Fig.~\ref{fig:chpt}:
\begin{align}
R_r(\tau) 
&= 1 + |\wt\tau| 
\left[\cA^{\rm LO}_{{\rm th}, r} + c_1 \frac{(\cA^{\rm LO}_{{\rm th},r})^2}{16 \pi M_\pi L}\right]
+ \frac{\wt\tau^2}2 \left(\cA^{\rm LO}_{{\rm th}, r}\right)^2 + \dots\,,
\label{eq:Rrexpect}
\end{align}
where
\begin{equation}
\wt \tau = \frac{\tau}{4 M_\pi^2 L^3}\,,
\label{eq:tildet}
\end{equation}
and the ellipsis indicates terms of higher order in $|\wt\tau|$ and $1/L$, 
NLO contributions to $\cA_{\thr,r}$ in the $|\wt\tau|$ term,
as well as excited state contributions.}
\end{quote}
In other words, we can focus on the terms that lead to factors of $\cA_{\thr,r}^{\rm LO}$, 
with a full NLO calculation of this amplitude not required.
Evaluating Fig.~\ref{fig:chpt}(b) gives the result for $\cA_{\thr,r}^{\rm LO}$,
while Fig.~\ref{fig:chpt}(c) gives the $c_1$ and $\tau^2$ terms,
which must have the dependence on $\cA_{\thr,r}^{\rm LO}$ given in Eq.~(\ref{eq:Rrexpect}).
We also note that one can check the calculation of Fig.~\ref{fig:chpt}(b) by evaluating
the amplitude $\cA_{\thr,r}^{\rm LO}$ directly in infinite volume.

Passing this test is necessary in order that a channel can be treated as physical,
but it is clearly not sufficient, since higher order terms in $\tau$ and in the threshold expansion
of Eq.~(\ref{eq:LuscherTH}) are not checked. In addition, the test considers only the
threshold state, while a complete test would also consider excited states.
Possible generalizations are discussed below in Sec.~\ref{sec:gen}.

\section{PQ\chpt\ calculation of $\pi\pi$ correlators}
\label{sec:calc}

We perform the test described in the previous section for the $\bm{20}$ and $\bm{15}$ irreps,
which are the two that are absent in the unquenched theory.
For comparison we also include the $\bm{84}$ irrep.
Examples of two-pion operators transforming in these irreps are
\begin{align}
\cO_{\bm{84}} &= \frac1{\sqrt2} \left( \wt\pi_{12} \wt\pi_{34} + \wt\pi_{14}\wt\pi_{32}\right)
\label{eq:O84}
\\
\cO_{\bm{20}} &= \frac1{\sqrt2} \left( \wt\pi_{12} \wt\pi_{34} - \wt\pi_{14}\wt\pi_{32}\right)
\label{eq:O20}
\\
\cO_{\bm{15}} &= \frac1{\sqrt3} \left( \wt\pi_{13} \wt\pi_{32} +  \wt\pi_{14} \wt\pi_{42}
+\wt \pi_{12} \wt \eta_4 \right)\,.
\label{eq:O15}
\end{align}
The normalization factors are chosen as described in the previous section,
i.e. such that the leading term in $R_\tau$ is unity.
In the notation of Ref.~\cite{Acharya:2019meo} these three irreps are labeled
$\alpha$, $\beta$ and $\gamma$, respectively.
The LO threshold amplitudes for these irreps
can be read off from the results for scattering lengths given in Ref.~\cite{Acharya:2019meo} 
(noting that the scattering length in that work is defined with opposite sign to ours),
yielding
\begin{equation}
\cA_{\thr,\bm{84}}^{\rm LO} = -\frac{M_\pi^2}{f^2}\,,\qquad
\cA_{\thr,\bm{20}}^{\rm LO} = \frac{M_\pi^2}{f^2}\,,\qquad
\cA_{\thr,\bm{15}}^{\rm LO} = \frac72 \frac{M_\pi^2}{f^2}\,.
\label{eq:Athres}
\end{equation}

\begin{figure}[tb]
\begin{center}
\vspace{-12pt}
\includegraphics[width=\textwidth]{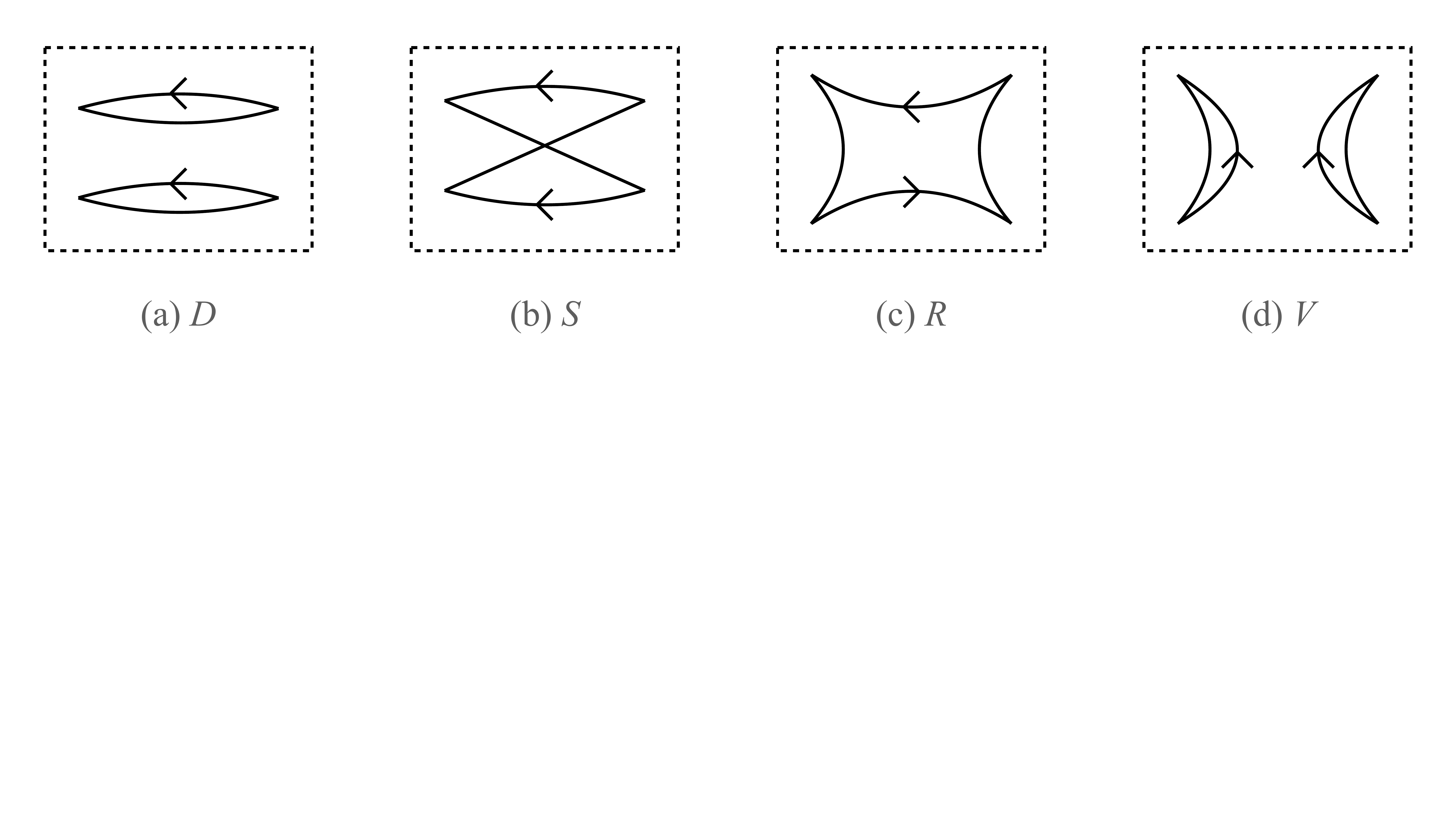}
\vspace{-2.5truein}
\caption{Types of Wick contraction that contribute to the two-pion correlators, $R_r(\tau)$.
Time runs horizontally. Solid lines represent quark propagators, while 
the dashed box is a reminder that the correlators are evaluated in finite spatial volumes.
\label{fig:contractions} 
}
\end{center}
\end{figure}

The implementation of the criterion C1 for the $\bm{84}$ and $\bm{20}$ irreps
can be done by combining results from Ref.~\cite{Hansen:2011mc}. In the notation of that work
\begin{equation}
R_{\bm{84}/\bm{20}}(\tau) = R_{\cD}(\tau) \pm R_{\cS}(\tau)\,,
\label{eq:RfromHS}
\end{equation}
where $R_{\cD}$ involves the double-loop (or ``direct'' or $D$ for short) 
Wick contraction in the numerator of the ratio,
while $R_{\cS}$ involves the single loop (``$S$'') contraction,
which are shown in Figs.~\ref{fig:contractions}(a) and (b), respectively.
In Refs.~\cite{Acharya:2017zje,Acharya:2019meo}, the latter contraction is labeled $C$ for
crossed.
The results for $D(\tau)$ and $S(\tau)$ at NLO in PQ\chpt\ are
given in Eqs.~(104) and (105) of Ref.~\cite{Hansen:2011mc}, including the effect of
discretization errors for Wilson-like fermions.
Setting all LECs associated with discretization errors to zero, and choosing 
$\tau>0$ as in Ref.~\cite{Hansen:2011mc}, the results read 
\begin{align}
R_{\cD}(\tau) &= 1 + \wt \tau \cD
+ \left(\frac{M_\pi^2}{ f^2}\right)^2 \left[ 
\wt\tau \frac{c_1}{16\pi M_\pi L} + \frac{\wt\tau^2}{2} \right] + \dots \,,
\label{eq:RDresHS}
\\
R_{\cS}(\tau) & = \wt \tau \cS + \dots \,.
\end{align}
Here $\cD$ and $\cS$ are the NLO PQ threshold amplitudes associated with the $D$ and $S$
contractions, and are given in Eqs.~(106) and (107), respectively, of Ref.~\cite{Hansen:2011mc}.
We only need their LO contributions, which are
\begin{equation}
\cD^{\rm LO} = 0\,, \qquad \cS^{\rm LO} = - \frac{M_\pi^2}{f^2}\,.
\label{eq:SDLOHS}
\end{equation}
Finally, we form the combinations given in Eq.~(\ref{eq:RfromHS}), obtaining
(again for $\tau>0$)
\begin{align}
R_{\bm{84}/\bm{20}}(\tau) &= 1 + \wt \tau (\cD \pm \cS) + \left(\frac{M_\pi^2}{ f^2}\right)^2 
\left[  \wt\tau \frac{c_1}{16\pi M_\pi L} + \frac{\wt\tau^2}{2} \right] + \dots \,,
\label{eq:R8420res}
\end{align}
Since $(\cD\pm\cS)^{\rm LO} = \mp M_\pi^2/f^2$, we see that the results for both
irreps satisfy the criterion C1, 
i.e. Eq.~(\ref{eq:Rrexpect}) with the LO amplitudes given in Eq.~(\ref{eq:Athres}).\footnote{%
This result in fact holds even if we keep all contributions from the LO LECS associated
with discretization errors.}
This is as expected for the physical $r=\bm{84}$ channel, 
while, for the $r=\bm{20}$ channel, it supports the proposal of 
Ref.~\cite{Acharya:2019meo} that one can apply the quantization condition to this
unphysical channel.

Now we turn to the $\bm{15}$ irrep, for which a new calculation is needed.
In terms of valence quark Wick contractions, one needs not only $D$ and $S$ contractions,
but also the ``rectangle'' or $R$ contraction 
shown in Fig.~\ref{fig:contractions}(c).
We find, using $\cO_{\bm{15}}$ from Eq.~(\ref{eq:O15}), that
\begin{equation}
R_{\bm{15}}(\tau) = R_{\cD}(\tau) - \frac12 R_{\cS}(\tau) + 3 R_{\cR}(\tau)\,,
\label{eq:R15decomp}
\end{equation}
in agreement with Ref.~\cite{Acharya:2019meo}.
Evaluating the diagrams in Fig.~\ref{fig:chpt}, we obtain (with $\wt\tau>0$)
\begin{equation}
R_{\bm{15}}(\tau) = 1 + \wt\tau \frac{7 M_\pi^2}{2 f^2}
+ \underbrace{\left[\frac{49}4 - 6- \frac{3}{4}\right]}_{11/2} \left(\frac{M_\pi^2}{f^2}\right)^2
\left[  \wt\tau \frac{c_1}{16\pi M_\pi L} + \frac{\wt\tau^2}{2} \right] + \dots \,,
\label{eq:R15res}
\end{equation}
where the ellipsis has the same meaning as in Eq.~(\ref{eq:Rrexpect}).
The calculation leading to Eq.~(\ref{eq:R15res}) is straightforward, 
as the $c_1 \wt \tau$ and $\wt \tau^2$ terms essentially
pick out contributions that are part of the square of the tree-level threshold amplitude.
The only subtlety is keeping track of which two-particle intermediate states contribute,
and of cancellations between contributions involving PG bosons and fermions.
We have broken down the contribution to the final term in Eq.~(\ref{eq:R15res}) into that from
the following intermediate states: $\pi_{13}\pi_{32}$,  $\pi_{14}\pi_{42}$ and $\pi_{12}\eta_4$
(leading to the unquenched result $49/4=(7/2)^2$),  $\omega_{15}\omega_{52}$ and
$\omega_{16}\omega_{62}$ (leading to the $-6$), and $\pi_{12} \phi_1$
(leading to the $-3/4$). We have obtained these results using both the straceless
form of $\Pi$ given in Eq.~(\ref{eq:Pi}) and an approach in which one projects out the stracefull
part by introducing a $\Phi_0^2$ term~\cite{Sharpe:2001fh}.

Equation~(\ref{eq:R15res}) is the main new result of this paper. It is manifestly
inconsistent with the form of Eq.~(\ref{eq:Rrexpect}), because the
coefficient of the $\wt\tau^2$ and $c_1\wt\tau$ contributions is not the square of the
LO threshold amplitude (which itself is correctly determined from the coefficient of $\wt\tau$).
Thus, the $r=\bm{15}$ channel fails our test: $R_{\bm{15}}(\tau)$ 
cannot be given by a single exponential at long times,
and the two-particle quantization condition cannot be used.

\section{Diagrammatic interpretation and alternative criteria}
\label{sec:gen}

\begin{figure}[tb]
\begin{center}
\vspace{-10pt}
\includegraphics[width=\textwidth]{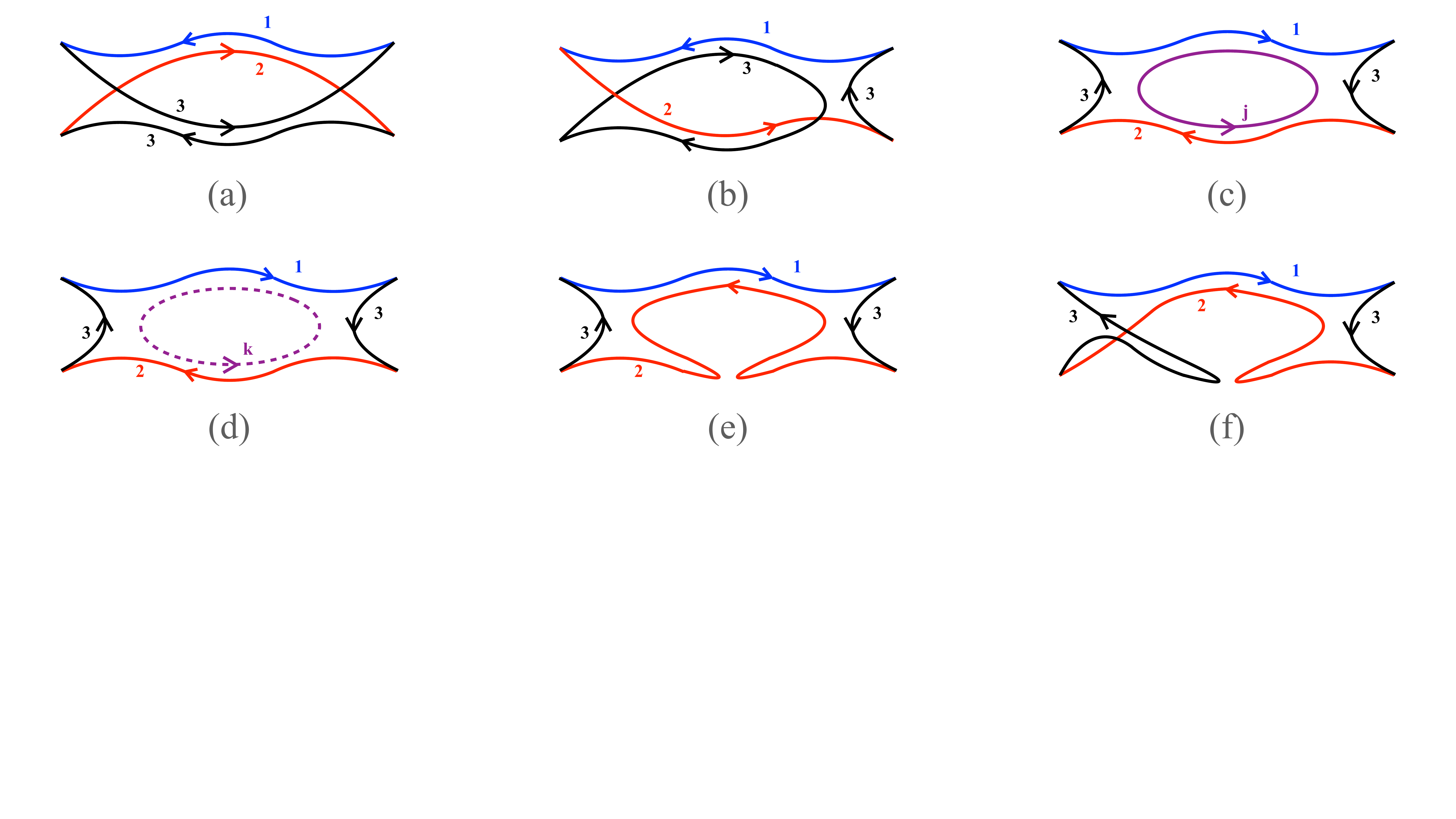}
\vspace{-1.9truein}
\caption{Classes of quark-line diagrams contributing to Fig.~\ref{fig:chpt}(c) for the
$\bm{15}$ irrep.
The diagrams trace the flavor indices of the quarks and antiquarks (or their ghost partners)
in the PQ\chpt\ diagrams. Flavors are indicated by colors and numerical labels, with
$j=1-4$ in (c) and $k=5-6$ in (d). Dashed lines indicate ghost-quark flavors.
In diagrams (e) and (f) the hairpin vertices implicitly contain an infinite sum of 
intermediate loops of sea quarks (with the valence and ghost loops canceling).
\label{fig:quarklines}
}
\end{center}
\end{figure}

To gain more understanding of why the $\bm{15}$ irrep does not satisfy our criterion, 
while the $\bm{20}$ does,
we consider the quark-line diagrams\footnote{%
For more discussion of quark-line diagrams see Ref.~\cite{Sharpe:2006pu}.}
associated with the corresponding contributions
to Fig.~\ref{fig:chpt}(c). 
These are shown for the $\bm{15}$ irrep in Fig.~\ref{fig:quarklines}.
The key observation is that, for $r=\bm{20}$, only diagrams of the type 
of Fig.~\ref{fig:quarklines}(a) appear, with the four lines having different flavors.
This difference arises simply because there can be no ``annihilation'' of quark flavors for
$r=\bm{20}$, which is why the $R$-type Wick contraction, Fig.~\ref{fig:contractions}(c), is not present.
This absence implies that there is no violation of unitarity in the $s$-channel in the one-loop
diagrams, and it is this form of unitarity (which we refer to as $s$-channel unitarity)
that the exponential time dependence
and the derivation of the quantization condition depend on.

Conversely, for $r=\bm{15}$, the appearance of $s$-channel loops involving either ghost
quarks [Fig.~\ref{fig:quarklines}(d)] or neutral PGBs with negative norms such as $\phi_0$
[Figs.~\ref{fig:quarklines}(e) and (f)] suggests that $s$-channel unitarity is violated.
We emphasize, however, that it is not sufficient to look at the diagrams alone,
as the $s$-channel-unitarity-violating contributions can cancel.
This indeed happens in the $I=0$ channel, for which there are both $R$ and $V$ Wick 
contractions [Figs.~\ref{fig:contractions}(c) and (d)], and thus PQ diagrams involving
ghost quarks and negative norm PGBs. Nevertheless, we know that these unphysical
contributions must cancel,
as this channel is physical.

Putting this together, the following diagrammatic version of the criterion seems highly plausible:
\begin{quote}
C2: {\em
A channel that is not equivalent to a physical correlator will not be described by the two-particle quantization condition if intermediate ghost quarks appear in (the quark-line diagrams of) s-channel two-particle loops.}
\end{quote}
This criterion correctly selects the $\bm{15}$ irrep as the single PQ 
nonsinglet $\pi\pi$ channel that is problematic.
We have not included negative norm PGBs in the statement  of C2 as it requires explicit calculations
to determine whether they are present, and we are aiming here for a simplified criterion.
We suspect, however, that nothing is lost by this
omission, since, in our experience, diagrams with intermediate
ghost quarks and negative norm PGBs come in tandem.
C2 is potentially stronger than our original criterion, C1, as it applies away from threshold.
We have, however, not found a general proof of C2, since this requires knowing that
the total contribution from certain classes of quark-line diagrams does not cancel,
and this can, in general, only be determined by an explicit calculation, leading us back 
to C1.

\begin{figure}[b]
\begin{center}
\vspace{-10pt}
\includegraphics[width=\textwidth]{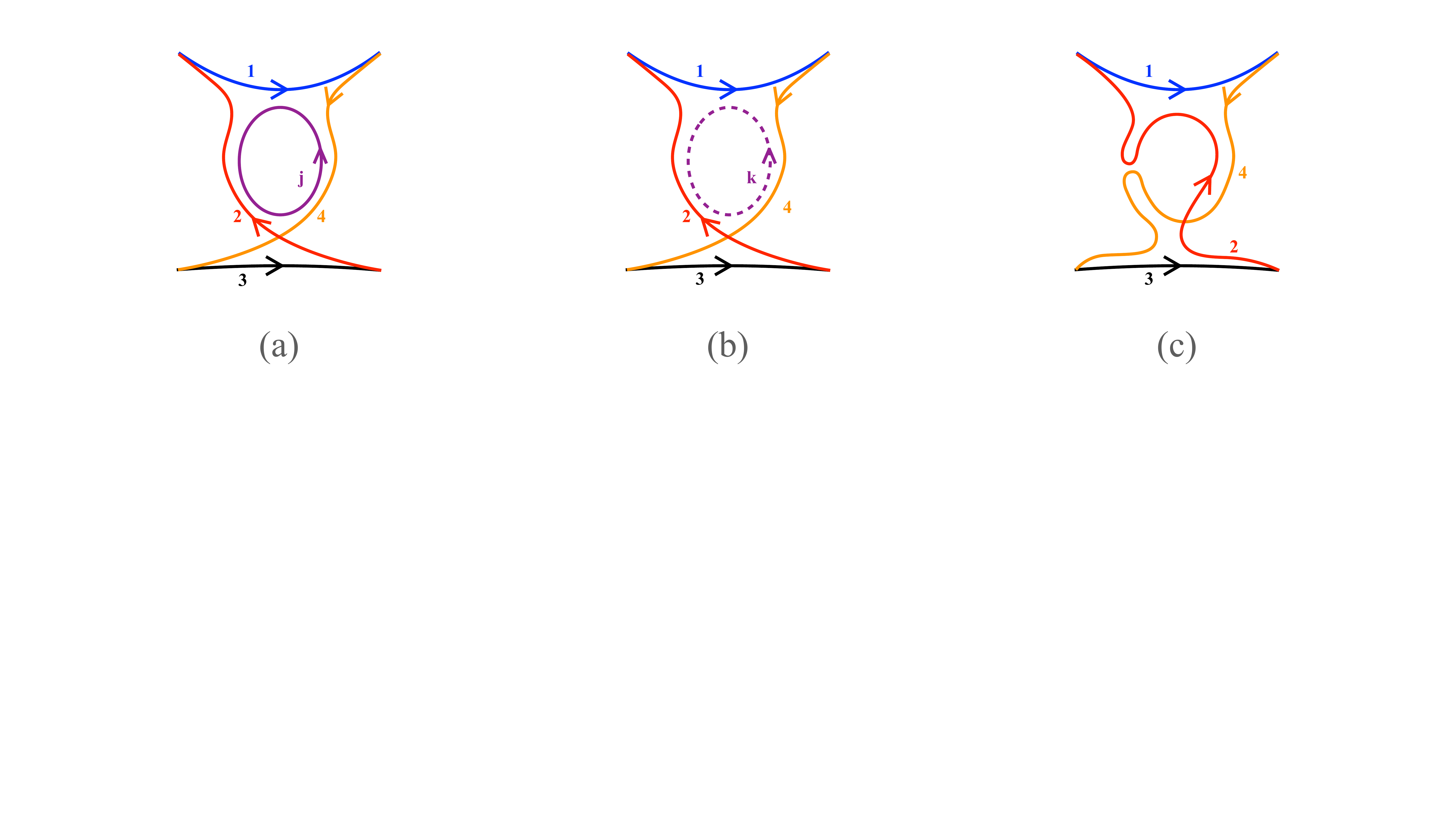}
\vspace{-2.3truein}
\caption{Examples of quark-line diagrams contributing to the $t$-channel loops for scattering
in the $\bm{20}$ irrep. Notation as in Fig.~\ref{fig:quarklines}. 
\label{fig:quarklines-t}
}
\end{center}
\end{figure}

These diagrammatic considerations lead to another version of the criterion.
To explain this, we need to distinguish between unitarity in the $s$, $t$ and $u$ channels.
As noted above, 
we refer to the standard form of unitarity for an amplitude $\cT(s,t,u)$ in the physical domain
($s \ge 4 M_\pi^2$ and $t,u \le 0$) as $s$-channel unitarity.
One can also apply unitarity in the $t$ channel by continuing the amplitude
to the appropriate kinematic domain, $t \ge 4 M_\pi^2$ and $u,s \le 0$, and
similarly for the $u$ channel.
An important point to keep in mind is that PQ amplitudes can satisfy unitarity in some channels
but not in others.
For example, the amplitude $\cA_{\bm{20}}$ satisfies $s$-channel unitarity, 
but violates $t$- and $u$-channel unitarity.
This is illustrated in Fig.~\ref{fig:quarklines-t}, where examples of quark line diagrams
contributing to $t$-channel loops are shown.
The presence of the diagrams (b) and (c) indicates the violation of $t$-channel unitarity.

Using this terminology, the third version of the criterion is
\begin{quote}
C3: {\em A channel is not described by the two-particle
quantization condition if the (infinite-volume) amplitude $\cA_r$ violates
$s$-channel unitarity.}\footnote{%
We stress that we are only considering energies below the first inelastic threshold, e.g.
below the four pion threshold for $\pi\pi$ scattering. Above this,
the two-particle quantization condition itself breaks down and the question becomes moot.}
\end{quote}
Again, this correctly picks out the $\bm{15}$ $\pi\pi$ irrep.
The simplest argument for C3 is that $s$-channel unitarity plays a crucial role in the
derivation of the two-particle quantization condition, as is particularly clear from the
derivation of Ref.~\cite{\KSS}.
C3 is stronger than the original form, C1, 
as the latter involves unitarity applied only infinitesimally above threshold,
while C3 applies for all energies up to the first inelastic threshold.

To study C3 in \chpt\ requires at least a NLO calculation, since loops are required for
the unitarity condition to have nontrivial content.
Indeed, from the NLO results for PQ $\pi\pi$ amplitudes in Ref.~\cite{Acharya:2017zje},
one can see that $s$-channel unitarity is violated by $\cA_{\bm{15}}^{\rm NLO}$.
In general, C3 is harder to implement than C1, as it
 requires a more extensive calculation in \chpt.

We end this section by discussing whether the criteria that
we have introduced are actually sufficient rather than necessary.  
We suspect that this is the case, although to demonstrate this would
require an all-orders analysis in \chpt, which is beyond the scope of this work.
The point is that higher order terms in the exponential in $\tau$, and in the threshold
expansion of the quantization condition, are built up by a sequence
of intermediate two-particle $s$-channel
loops, as is clear, respectively, from the calculations of Ref.~\cite{Hansen:2011mc}
and the derivation of Ref.~\cite{\KSS}.
The corresponding quark-line diagrams
for any irrep that passes the criteria will only involve
generalizations of Fig.~\ref{fig:quarklines}(a), in which quarks or antiquarks are
repeatedly exchanged. 
In particular, there can be no annihilation diagrams at any order in \chpt.
The unphysical contributions in $t$ and $u$ channels,
such as those in Fig.~\ref{fig:quarklines-t}, will be present, but will appear in all vertices,
allowing exponentiation.

\section{Applications and implications for previous work}
\label{sec:implications}

We first summarize our results for the $\pi\pi$ system.
In the PQ theory there are  
three flavor nonsinglet channels, of which one (the $\bm{84}$)
is equivalent to a physical channel (with $I=2$), which thus
automatically satisfies our criteria.
Of the unphysical channels, one satisfies our criteria ($\bm{20}$), while
the other does not ($\bm{15}$).

What are the implications for the application presented in Ref.~\cite{Acharya:2019meo}?
In that work, the quantization condition was first applied to LQCD results in the $\bm{84}$ and
$\bm{20}$ channels, and this application is thus not affected by the problem noted here.
The results obtained were then used to predict the 
finite-volume energy spectrum in the $\bm{15}$ and $I=0$ channels, 
while our analysis shows that there is no well-defined spectrum in the $\bm{15}$ channel.

We next turn to the $K\pi$ system, which has been analyzed in this context in
Refs.~\cite{KalmahalliGuruswamy:2020uxi,Guruswamy:2021nph}.
Since there are now three light quarks, the isospin symmetry generalizes from 
$\SU(2)$ in the physical theory to $\SU(3)$ in the PQ version, which is the light
quark subgroup of the graded $\SU(4|1)$ symmetry.
This $\SU(3)$ is exact (assuming $m_u=m_d$),
and should not be confused with the approximate $\SU(3)$ in QCD involving the strange quark.
There are two physical channels, with $I=3/2$ and $1/2$, while there are three PQ channels,
corresponding to the $\SU(3)$ decomposition 
$\bm{8} \times \bm{3}= \bm{15} + \overline{\bm{6}} + \bm{3}$. 
These three channels are labeled $\alpha$, $\beta$, and $\gamma$, respectively, 
in  Ref.~\cite{KalmahalliGuruswamy:2020uxi}.
Of these, only the $\overline{\bm{6}}$ is unphysical, with the $\bm{15}$ being equivalent to
the physical $I=3/2$ channel, and the $\bm{3}$ to that with $I=1/2$.

Applying the criterion C2 to the $\overline{\bm{6}}$ channel, we find that it passes our test, because
it involves Wick contractions only of types $D$ and $S$, 
and is thus completely analogous to the $\bm{20}$ irrep for $\pi\pi$ scattering.
Therefore, the numerical application presented in
Ref.~\cite{Guruswamy:2021nph}, which uses the $\bm{15}$ and $\overline{\bm{6}}$
channels, does not suffer from the problem described here.

The final example we consider involves two kaons, for which
the PQ extension has not been previously analyzed.
For the $KK$ system, the PQ extension involves treating the two strange antiquarks as different,
while the two light quarks can both be physical.
Thus the graded symmetry group becomes $\SU(2)\times\SU(2|1)$,
where the $\SU(2)$ is physical isospin. 
Thus PQ channels can be labeled by physical isospin, $I$, as well as 
``strange isospin'', $I_s$, with both taking on the values $0$ or $1$.
For $s$-wave scattering, the overall symmetry implies that there are two
allowed channels, with $(I,I_s)=(1,1)$ or $(0,0)$, which are, respectively,
symmetric and antisymmetric under quark (or antiquark) exchange . Only the former is physical.
Again, the sole unphysical channel, $(0,0)$, satisfies our criterion C2, since only
contractions of type $D$ and $S$ are allowed as there is no annihilation in the $s$ channel.
Thus the quantization condition could be applied to both channels.

\section{Conclusions}
\label{sec:conc}

This paper has demonstrated that the violation of unitarity in PQ theories 
at best restricts the application
of the two-particle quantization condition to a subset of the unphysical two-particle channels.
This has been demonstrated using PQ\chpt, and our initial criteria C1 is stated within this
effective field theory (EFT) framework. In this regard, we stress that the foundations of PQ\chpt\
as an effective theory for PQQCD are almost as strong as for QCD~\cite{Bernard:2013kwa}.
However, we have argued that C1 can be replaced
by  simpler, diagrammatic extension of the criterion, C2.
This alternative version can be used irrespective of the existence of an EFT.

What is perhaps most surprising is that there may be unphysical two-particle channels
in which the quantization condition {\em can be used}, because the derivation of the latter relies
on unitarity, which is certainly violated in the PQ theory.
The key observation here is that what matters for the derivation is 
that the corresponding scattering amplitude is unitary in the $s$ channel, while
$t$- and $u$-channel unitarity is not required. Thus for channels where
the amplitude manifests its unphysicality only in the latter two channels, the
problem that we raise does not occur.

We stress that, strictly speaking, we have not demonstrated this positive result,
but rather only shown that one possible problem does not arise.
However, we think it likely that one could extend the argument given here to all orders
in the EFT, and that, for channels that satisfy our simple criteria, the use of the
quantization condition is justified.

We also note that all is not lost in the channels that do not satisfy our criteria.
One can still calculate the corresponding finite-volume correlators using PQ\chpt,
fit the results of a lattice simulation to the predicted form, and in this way
determine the parameters of the EFT. This is the approach suggested in Ref.~\cite{Hansen:2011mc}.
We stress, however, that our results show that one cannot bypass the problem
by determining the energy shift solely from the term in the correlator ratio that is linear in $\tau$, 
and inserting the result into the two-particle quantization condition, 
because the leading $c_1/L$ correction in that condition does not hold in PQ channels
that do not satisfy our criteria.

Applying our criteria to the $\pi\pi$ and $K\pi$ systems, we find that the
two previous applications which use results from simulations of LQCD are in 
unphysical channels that are {\em not} affected by the issue that we raise. 
Thus, the practical impact of our observation
is to limit certain future applications of the approach of Refs.~\cite{Acharya:2019meo,KalmahalliGuruswamy:2020uxi}.

One way of understanding our central result is that, in a PQ theory, even
if one starts with an operator that is composed of sea and valence quarks alone,
in general there will be mixing with operators that include ghost quarks.
In the calculation of Sec.~\ref{sec:calc}, this mixing was with operators of the
form $\omega_{1k}\omega_{k2}$ and $\pi_{12} \phi_1$, both of which transform in the
same $\bm{15}$ irrep of the $\SU(4)$ quark flavor group. 
One might hope that, by taking an appropriate linear combination of
these operators, one could find a channel for which one could use the quantization condition.
This amounts to classifying operators under the full $\SU(4|2)$ group, a task that we have
not undertaken. We do not, however, expect this approach to be fruitful, since intermediate
states will inevitably involve unphysical particles.
For the same reason, we expect that the $\SU(4)$ singlet channel, where there can
be mixing with several unphysical channels (including $\phi_0^2$), cannot be analyzed
using the quantization condition.

A final comment concerns the foundational work of
Bernard and Golterman~\cite{Bernard:2013kwa}.
They construct the transfer matrix of (latticized) PQQCD, showing that, while it is not hermitian,
it has a bounded spectrum. The import is that one expects correlation functions to still be
given by a sum of exponentials as in Eq.~(\ref{eq:2pt}), except the coefficients $c_n$
need not be positive. Furthermore, the energies $E_n$, while having real parts
that are bounded from below, could be complex, although Ref.~\cite{Bernard:2013kwa} argues
that this is not the case for the lightest excitations in PQ\chpt.
Thus it could be that one can develop a generalization of the two-particle
quantization condition that  takes into account the possibility of negative $c_n$,
and the concomitant cancellations. The lack of a simple exponential fall off that we have found
could then be due to a cancellation between terms falling with similar exponents
but opposite coefficients.
The failure of the standard L\"uscher threshold expansion could be due to the need to
use an extension of the two-particle formalism.
We do not know if such an approach will be fruitful, but it may be worth considering.

\section*{Acknowledgments}

We thank Maarten Golterman, Max Hansen, Ulf Mei{\ss}ner and Fernando Romero-L\'opez for very helpful comments and suggestions. 
This work is supported in part by U.S. Department of Energy Award No. DE-SC0011637.

\bibliography{ref}

\end{document}